\documentclass[12pt]{article}
\usepackage{graphicx,amssymb}
\usepackage{bm,enumerate}

\newcommand{\beq}{\begin{equation}}
\newcommand{\eeq}{\end{equation}}

\begin{document}

\title{On the Security of the Y-00 \hspace{0.15cm}($\alpha\eta$)\hspace{0.15cm} Direct Encryption Protocol}

\author{Ranjith Nair$^\dag$ \& Horace P.~Yuen\\Department of Electrical Engineering and Computer Science, \\
Northwestern University, Evanston, IL, 60208, \\
$^\dag$Email: nair@eecs.northwestern.edu} \maketitle

\begin{abstract}
We review the current status of the $\alpha\eta$ direct encryption
protocol. After describing $\alpha\eta$, we summarize the main
security claims made on it. We then describe recent attacks
developed against it in the literature, and suggest security
enhancements and future research directions based on our results.
\end{abstract}

\section{Introduction}
This article summarizes a poster presentation at QCMC 2006 on the
security of the $\alpha\eta$ protocol. The $\alpha\eta$ protocol
\cite{yuen03,barbosa03,nair06,corndorf05,yuen05} was developed as
an efficient (the `$\eta$' in $\alpha\eta$) quantum encryption
protocol using coherent states (`$\alpha$'). Its objective being
direct data \emph{encryption}, it is inappropriate to compare it
with quantum cryptographic protocols for \emph{key generation},
such as BB84, continuous-variable QKD, and entanglement-based QKD
protocols. First of all, $\alpha\eta$ uses a pre-shared secret key
(typically a few thousand bits long) that is not assumed in key
generation protocols (except for a short authentication key).
Secondly, the criterion of success of an encryption protocol is
not so stringent as in a key generation protocol, where one
ideally desires to distill bits that are nearly random to Eve. For
the first reason given above, it is also inappropriate to compare
$\alpha\eta$ to a composite protocol in which , e.g., BB84 is used
to generate nearly random keys which are subsequently used for
data encryption through, e.g., one-time pad. On the other hand,
from a cryptographic standpoint, one can make a fair comparison
between $\alpha\eta$ and a standard classical encryption protocol
like one-time pad or AES since the cryptographic objective is the
same in both cases. Unfortunately, to our knowledge, there is no
universally agreed upon security criterion for standard encryption
which can be calculated for any meaningful standard cipher
(excluding one-time pad). Thus, security claims are usually made
given certain unproved assumptions and in some cases these
assumptions have only sociological support. Given this situation,
we will take care to state all the assumptions made for our claims
in the rest of this article.

\section{The $\alpha\eta$ cryptosystem}
We now describe the steps of operation of an $\alpha\eta$
cryptosystem as depicted in Fig.~1:
\begin{enumerate}[(1)]

\item
Alice and Bob share a secret key $\mathbf{K}$.

\item
Using a \emph{key expansion function} $ENC(\cdot)$, e.g., a linear
feedback shift register or AES in stream cipher mode, the seed key
$\mathbf{K}$ is expanded into a running key sequence that is
chopped into $n$ blocks: $\mathbf{K}_{Mn}=ENC(\mathbf{K})=(K_1,
\ldots , K_{mn})$. Here, $m=\log_2(M)$, so that $Z_i \equiv
(K_{(i-1)m +1}, \ldots, K_{im})$ can take $M$ values. The $Z_i$
constitute the \emph{keystream}.

\item
%The encrypted state $e_{\mathbf{K}_s}(\mathbf{X}_n)$ of
%Eq.(\ref{quantcipher})is defined as follows.
For each bit $X_i$ of
the plaintext sequence $\mathbf{X}_n = (X_1, \ldots, X_n)$, Alice
transmits the \emph{coherent state}
\begin{equation} \label{state}
|\psi(X_i,Z_i)\rangle=|\alpha e^{i\theta(X_i,Z_i)}\rangle.
\end{equation}
Here, $\alpha \in \mathbb{R}$ and $\theta(X_i,Z_i)$ takes values
in the set $\{0,\pi/M,\ldots,(2M-1)\pi/M\}$. The function $\theta$
taking the data bit and keystream symbol to the actual angle on
the coherent state circle is called the \emph{mapper}. In this
article, we assume that $\theta(X_i,Z_i)=[Z_i/M+(X_i\oplus
Pol(Z_i))]\pi$. $Pol(Z_i)= 0$ or $1$ according to whether $Z_i$ is
even or odd. Thus $K_i$ can be thought of as choosing a `basis'
with the states representing bits $0$ and $1$ as its end points.

\item
In order to decrypt, Bob runs an identical ENC function on his
copy of the seed key. For each $i$, knowing $Z_i$, he makes a
quantum measurement to discriminate the two states
$|\psi(0,Z_i)\rangle$ and $|\psi(1,Z_i)\rangle$ and recover the
input bit.
\end{enumerate}

\begin{figure*} [htbp]
\begin{center}
\rotatebox{-90} {
\includegraphics[scale=0.5]{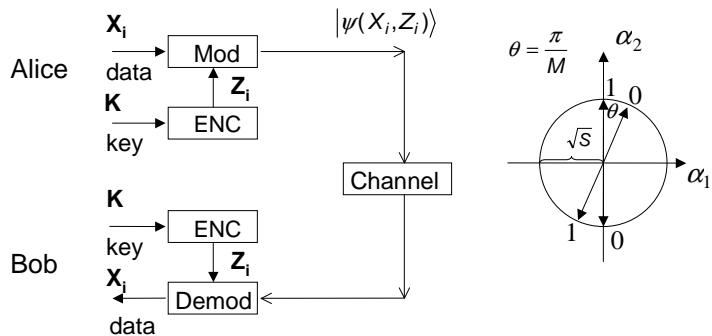}}
\caption{Left -- Overall schematic of the $\alpha\eta$ encryption
system.
 Right -- Depiction of two of $M$ bases with interleaved logical
bit mappings.}
\end{center}
\end{figure*}

\section{Security Claims}

We list in this section our theoretical claims regarding
$\alpha\eta$, leaving a discussion of some attacks on it to the
next section.

\begin{enumerate}

\item \emph{Random Cipher Character}: First, we claim \cite{nair06,corndorf05,yuen05} that the fundamental
performance of $\alpha\eta$ is equivalent to that of a
corresponding classical \emph{random} cipher when Eve makes
individual identical heterodyne or phase measurements on each
optical qumode. A random cipher differs from a non-random one in
associating more than one ciphertext to every plaintext-key pair.
For known-plaintext attack on the key, we have defined in
\cite{nair06} a parameter $\Gamma$ that is a measure of the number
of running keys that can be associated to a given
plaintext-ciphertext symbol pair, which we expect to be, at least
qualitatively, relevant to security. Under heterodyne attack, we
estimate $\Gamma \sim M/(\pi \sqrt{S})$ for signal energy $S$.
This number works out to around 3 for the typical parameters
$M\sim 2000, S \sim 40000$ used in \cite{corndorf05}. Further
details on the above, including why random ciphers are
theoretically interesting from a security standpoint, can be found
in \cite{nair06}.

\item \emph{Assisted Brute-Force Search Complexity}: One may
easily see that a heterodyne measurement by Eve on each qumode $i$
gives her partial information on the keystream symbol $Z_i$,
especially the most significant bits of $Z_i$ for the mapping
scheme of the previous section. In our so-called `wedge
approximation' \cite{nair06}, she may thus tabulate the possible
keystream sequences given her measurement for each $i$. We define
in \cite{nair06} an \emph{assisted brute-force search} attack on
the key to be an attack where Eve exhaustively checks (using any
algorithm) for a seed key that is compatible with one of the
keystream combinations. The factor by which her complexity
increases we call the assisted brute-force search complexity. For
example, when the ENC box is an LFSR, we show that it equals
$C=\Gamma^{|K|/\log_2{M}}$.

\item \emph{Ciphertext-Only Attack Security with DSR}: In
\cite{yuen06}, we detail a technique called Deliberate Signal
Randomization (DSR) involving a randomization of the state in
Eq.~(1) by Alice before transmission with the purpose of rendering
the seed key inaccessible to Eve in a ciphertext-only attack,
i.e., an attack where each data bit is independently completely
random. We show therein that DSR may be done in principle, namely
in the limit $S,M \rightarrow \infty, M/\sqrt{S} = \pi\Gamma$, at
the same time preserving the $\Gamma$ that Eve sees for
mode-by-mode measurements and increasing Bob's decoding error
probability  using the same decoding apparatus by an arbitrarily
small amount. This result demonstrates that $\alpha\eta$ can in
principle approach similar security under ciphertext-only attacks
as that obtained from standard stream ciphers \cite{yuen05}, even
if joint quantum attacks are made.

\end{enumerate}

\section{Recent Attacks on $\alpha\eta$}
In this section, we comment on some recent attacks on $\alpha\eta$
made by the Donnet group in \cite{donnet06} and by ourselves.
Earlier attacks by Lo and Ko and by the Nishioka group have been
addressed by us in detail in the papers \cite{yuen05} and
\cite{nair06} respectively.

\subsection{Correlation Attacks}
Donnet et al describe in \cite{donnet06} an attack based on the
viewpoint that the seed key is presented to Eve making heterodyne
measurements in a coded, i.e., redundant, form with noise on top.
For the LFSR case, a linear decoding algorithm may thus be
employed to retrieve the seed key from observations. While the
efficacy of such an attack for $|K|=32$ has been demonstrated, we
have commented in \cite{yuen06} that the linear decoding approach
is exponentially complex with respect to the key size and the
number of LFSR taps, both of which can be increased to make such
attacks impossibly complex. We also mentioned some security
measures that break the linear code structure and render linear
decoding algorithms ineffective. We also showed how $\alpha\eta$
with an ENC using a parallel configuration of AES boxes can be
used to provide more security than a single AES box.

\subsection{Joint Attack on $\alpha\eta$: Preliminary Results}
All the preceding results, except the one on DSR, are concerned
with attacks where Eve makes identical mode-by-mode quantum
measurements. Although impractical at present, her most general
attack is a joint measurement of the entire qumode sequence. For
the case of known-plaintext attack on the key, with the
conservative assumption that Eve is given a full copy of the
transmitted quantum state, the relevant quantity is her average
error probability $\overline{P}_e$ of discriminating the $|K|$
states given by products of states of the form of Eq.~(1) for a
given plaintext sequence $\mathbf{x}$. In \cite{nairthesis}, we
developed a new general technique of upper-bounding
$\overline{P}_e$. Applying it to $\alpha\eta$ with LFSR as the ENC
box, and for the parameters mentioned above and $|K| = 4000$ bits,
we find that Eve's error probability becomes completely negligible
for data length $n$ in the range of 10-100 Mbits. Since this is
based on an upper bound, the system could in fact be insecure for
smaller $n$. This result is not too surprising, as non-random
`nondegenerate' ciphers are also broken at their nondegeneracy
distance \cite{nair06,yuen05}, which is believed to be quite
small.

\section{Conclusion}
The insecurity of the bare $\alpha\eta$ under joint attack implies
that the random cipher character of $\alpha\eta$ is not sufficient
to provide a significant level of information-theoretic security.
However, the system would still have great practical value if it
possessed a high, e.g., exponential level of complexity-based
security. Thus, it seems that the study of complexity-based
security of random ciphers, and of quantitative security measures
in general, is important.

\section{Acknowledgement}

We would like to thank E.~Corndorf, G.~Kanter, P.~Kumar, and
T.~Eguchi for many useful discussions on the topics of this paper,
which was supported by DARPA under grant F 30602-01-2-0528 and
AFOSR under grant F A 9550-06-1-0452.

\end{document}